\documentstyle [12pt] {article}
\begin {document}
\large
\begin {center}
Defect Mass in Gravitational Field and Red Shift of Atomic and Nuclear
Radiation Spectra \\
\vspace{1cm}
Kh.M. Beshtoev\\
\vspace{1cm}
Joint Instit. for Nucl. Reser., Joliot Curie 6, 141980 Dubna,\\
Moscow region, Russia\\

\end {center}

\par
Abstract \\
\par
It is shown, that radiation spectrum of atoms (or nuclei) in
the gravitational field has a red shift since the effective mass of
radiating electrons (or nucleons) changes in this field. This red
shift is equal to the red shift of radiation spectrum in the gravitational
field measured in existence experiments. The same shift must arise when
the photon (or $ \gamma $ quantum) is passing through the gravitational
field if it participates in gravitational interactions (photon has no rest
mass). The absence of the double effect in the experiments, probably,
means that photons (or $ \gamma $ quanta) are passing through the
gravitational field without interactions.\\


\section {Introduction}
\par In works [1-3] the question on interpretation of the results of
measurements of the radiation spectrum of the gravitational field [4] was
discussed.  From the general point of view shift of the radiation spectrum
can be caused by:
\par
a) With changing of the radiating spectrum because of influence of the gravitational
field on the characteristics of the radiating particle;
\par
b) With changing of the photons (or $ \gamma $ - quanta) spectrum while their
passing through the gravitational field for their interactions;
\par
c) With the contribution of these both cases.

\section {\bf Defect of Mass and Red Shift of Spectrum
Radiations in Gravitational Field}

Influence of the gravitational field on atomic radiations can occur:
\par
1) Because of influence of the gravitational field on the radiating electron,
 rotating around a nucleus. In this case we must take into account the
contribution of the change gravitational field $ \varphi (r) $ on atomic
distances $a $:
$$
\Delta \varphi (r) = \varphi (r + a) - \varphi (r) \simeq
a \frac {\partial \varphi} {\partial r} \mid_r.
\eqno (1)
$$
It is obvious, that in weak gravitational fields this contribution can
be neglected.

\par
2) Because of the change of effective mass of radiating electron
(or nucleon) in the gravitational field $ \varphi (r) $ in point $r $.
A free electron (or nucleon) has mass $m $.
\par
It is well known, that electron connected in atom loses a part of mass
$ \Delta m $ which is equal to the energy of connection $ \Delta E $
(defect of masses). The same situation takes place, in much more degree,
in strong interactions, i.e. nucleus with nuclear number $A $ consisting
of $Z $ protons and $N $ neutrons has defect of mass $ \Delta M $, which
is equal to the energy of connection  of protons and neutrons $E _ {int} $:
$ \Delta M = E _ {int} $.
\par
Similarly to electromagnetic and strong interactions the gravitational
interaction, which is an attracted one, will cause a defect of masses,
determined by the gravitational potential $ \varphi (r) $ in
the point $r $ where the particle (electron or nucleon) is located:
$$
E _ {int} = m \varphi (r), \qquad \varphi (r) = - \frac {M G} {r},
\eqno (2)
$$
where $G $- gravitational constant, $M $ - weight of
the attracting system (Earth), and then
$$
\Delta m = \mid E _ {int} \mid = m \mid \frac {\varphi (r)} {c^2} \mid.
$$
\par
The difference of the gravitational interaction from electromagnetic
and strong ones consists in the absence of discrete states and also in
impossibility of energy loss while formation of the connected states
through this interaction (in  electromagnetic interactions it occurs
through radiation of photons and in strong interactions - through
radiation of hadrons). While formation of the connected states in
the gravitational interactions there is a mechanical loss of energy
(i.e. through strokes, and in terrestrial experiments with the help of
an expense of energy, which compensates this defect mass).
\par
So, in gravitational field $ \varphi (r) $ the electron
( or nucleon) effective mass $m_ {eff} $  is
$$
m _ {eff} = m (1 + \frac {\varphi (r)} {c^2}),
\eqno (3)
$$
i.e. it decreases by value $m \mid \frac {\varphi (r)} {c^2} \mid $.
Then a spectrum of electron radiation [5] in the gravitational field has
the following form:
$$
E = \frac {\alpha^2 m _ {eff} c^2} {2} \frac {Z^2} {n^2} [1 + \frac {\alpha^2
Z^2} {n} [\frac {1} {(j + 1/2)} - \frac {3} {4n}] +...],
\eqno (4)
$$
where
$$
\alpha = \frac {e^2} {4 \pi \hbar c}; \qquad n ' = 0, 1, 2...;
$$
$$
j = \frac {1} {2}, \frac {3} {2}, \frac {5} {2}...; \qquad
n = n ' + j + \frac {1} {2} = 1, 2, 3... \qquad ,
$$
and is displaced in the red side on the value  $ \Delta E $ (we suppose,
that in $E $  all thin effects connected to the other interactions,
except for gravitational ones are taken into account):
$$
\frac {\Delta E} {E} = \frac {\varphi (r)} {c^2},
\eqno (5)
$$
Or
$$
\frac {\Delta \nu} {\nu} = \frac {\varphi (r)} {c^2}.
\eqno (5 ')
$$
\par
In work [6] the red shift caused by the difference of gravitational
potentials on the surface of the Sun and the Earth was measured:
$$
\frac {\Delta \nu} {\nu} = \frac {\varphi_{sun} - \varphi_{earth}} {c^2},
$$
and there was obtained
$$
\frac {(\Delta \nu)_{exp}} {(\Delta \nu)_{theor}} = 1,01 \pm 0,06.
$$
\par
Energy levels of nuclei [7], as well as of atoms, probably, proportional
to the mass of radiating nucleon, therefore nuclear levels also will
be displaced in the gravitational field according to the formulae (3)
and (5) (it is interesting to note, that if the energy levels of nuclei
were back proportional to masses, the violet shift would occur).
\par
From (5) we see, that in point $r_1 $ in terrestrial gravitational
potential $ \varphi (r_1) $ the level schift is
$$
\frac {\Delta_1 E} {E} = \frac {\varphi (r_1)} {c^2},
\eqno (6)
$$
And in point $r_2 $ in the terrestrial gravitational potential
$ \varphi (r_2) $ the level shift is
$$
\frac {\Delta_2 E} {E} = \frac {\varphi (r_2)} {c^2},
\eqno (7)
$$
then the difference of levels in these two points is ($E = h \nu $)
$$
\frac {\Delta _ {12} E} {E} \equiv \frac {\Delta_{12} \nu} {\nu}
= \frac {(\varphi (r_1) - \varphi (r_2))} {c^2} =
\frac {\Delta \varphi} {c^2}.
\eqno (8)
$$
\par
The experimental results obtained in [4] have shown that in the
gravitational field there is a red displacement, by the same value
$ \Delta E $, determined by expression (8):
$$
\frac {(\Delta \nu) _ {exp}} {(\Delta \nu) _ {theor}} = 1,05 \pm 0,10
$$
And
$$
\frac {\Delta V} {2 c} = (0,9990 \pm 0,0076) \frac {\Delta \varphi} {c^2}.
$$
\par
Then, obviously, there does not remain any contribution, which is possible
due to the photon (or $ \gamma $ - quantum) interaction with
the gravitational field. We shall discuss the influence of the gravitational
fields on photon (or $ \gamma $ - quanta) spectra because of their
importance for the general relativity theory. Indeed, if photons
(or $ \gamma $ - quanta) pass through the gravitational field without
any interaction, in analogy with photon in the electrical field,
then the deflection of the light beam passing near the Sun is possible
to explain only its refraction in the Sun atmosphere.
Let us consider this question.

\section {\bf Change of the Spectrum of Photons (or $ \gamma $ - Quanta)
due to Their Interaction with the Gravitational field}

If the photon mass (further, in this section, we shall mention only photons
having in view that $ \gamma $ - quanta behave similarly) is determined
by the following expression [8] (see also references in [1-3]):
$$
m _ {ph} = \frac {E _ {ph}} {c^2}
\eqno (9)
$$
(it is necessary to note, that the massive particles interact in the
gravitational fields through rest masses $m $ but not  through
$m ' = \frac {E} {c^2} =m \gamma $), then while its movement
in the gravitational field because of the variable of this field,
there should be an interaction. In early interpretation (see references
in [3]) it was supposed that the red shift of the photon spectrum occurs
in the gravitational field because of such interaction. Then the photon
mass will  vary according to the standard formula:
$$
\Delta m ' _ {ph} = m _ {ph} \frac {\Delta \varphi} {c^2}.
\eqno (10)
$$
\par
It is obvious that light velocity  $c$ will depend on the gravitational
field and $c'(r) $ will have the following form:
$$
c ' (r) = \frac {c} {(1 - \frac {\varphi (r)} {c^2})}
\simeq c (1 + \frac {\varphi (r)} {c^2}),
\eqno (11)
$$
i.e. in the gravitational field the photon velocity will decrease and
the stronger is the field in point $r $  the less is the photon velocity.
\par
If the photon is moving from the point $r_1 $ to the point $r_2 $,
the velocity of the light will vary and, accordingly, the spectrum will
also vary. Then the frequency of photons changes according to the following
expression:
$$
\frac {\Delta \nu} {\nu} = \frac {(\varphi (r_1) -
\varphi (r_2))} {c^2}
\eqno (12)
$$
(we suppose that in the given point the standard ratio is fulfilled between
the photon characteristics, taking into account the variating of the light
velocity).
\par
With the physical point of view obviously that -- if the photon interacts
in the gravitational field then its velocity, frequency change and it
is  deflected and if the photon does not interact in the gravitational
field then  its  velocity, frequency does not change and it is not deflected.
\par
As it is already mentioned above, the experimental results obtained
in [4, 6] have shown that only the gravitational effect caused by
the defect of mass in gravitational field is observed, and the effect
caused by the photon interactions in the gravitational field is not
observed (in case if the photon interacts with the gravitational field,
the double effect should be observed in the experiments).  It is clear,
that this question requires to be studied.
\par
It is well known, that only massive bodies and particles participate
in the Newton theory of gravitation (i.e. body and particle having rest
mass).  Since the photons have no rest mass, the usage of the mass
$m_{ph}$ obtained in the formula (9) is a hypothesis to be check of.
The check has shown (see above) that, probably, there are no photons
(or $ \gamma $ - quanta) red shift when they are passing through
the gravitational field. It is clear since they have no rest masses.
In this case there is a question: How there can the deflection of photons
appear in the gravitational field, if they do not participate in these
interactions? It is clear, that this problem requires a solution in
the experimental aspect. Let's note, that the given question was discussed
in work [9] (see also references in [9]).

\section {Conclusion}

It was shown, that radiation spectrum of atoms (or nuclei) in
the gravitational field has a red shift since the effective mass
of radiating electrons (or nucleons) changes in this field. This red shift
is equal to the red shift of radiation spectrum in the gravitational
field measured in existence experiments. The same shift must arise when
the photon (or $ \gamma $ quantum) is passing through the gravitational
field if it participates in gravitational interactions (photon has no
rest mass). The absence of the double effect in the experiments, probably,
means that photons (or $ \gamma $ quanta) are passing through
the gravitational field without interactions.

\par
References
\par
\noindent
1. V.N. Strelt'sov, JINR Communic. P2-96-435, Dubna, 1996;
\par
JINR Communic. P2-98-435, Dubna, 1998.
\par
\noindent
2. V.V. Okorokov, ITEP preprint N 27, Moscow, 1998.
\par
\noindent
3. L.B. Okun, K.G. Selivanov, V.L. Telegdi, UFN (Russian Journ.), 1999, v.69,
\par
p.1140.
\par
\noindent
4. R.V. Pound, G.A. Repka, Phys. Rev. Let., 1960, v.4, p.337;
\par
R.V. Pound, J.L. Snider, Phys. Rev., 1965, v.140, p.788.
\par
\noindent
5. S.S. Schweber, An Introduction to Relat. Quantum Field Theory,
\par
Row-Peterson and Co., N. Y., 1961.
\par
\noindent
6. J.L. Snider, Phys. Rev. Let., 1972, v.28, p.853.
\par
\noindent
7. M. Preston, Physics of  Nuclei, M., Mir, 1961;
\par
O. Bohr and B. Mottelson, Nuclear Structure, v.1, M.,
\par
Mir, 1971.
\par
\noindent
8. A. Einstein, Ann. Phys. (Leipzig) 1916, v.49, p.769;
\par
L.D. Landau, E.M. Lifshits, Field Theory, M., Nauka, 1988,
\par
p.324.
\par
\noindent
9. P. Marmet and C. Couture, Physics Essays, 1999, v.12, p.162.

\end {document}